\documentclass[12pt]{article}

\usepackage{graphicx}
\usepackage{color}
\usepackage{epsfig}
\usepackage{latexsym}
\usepackage{bm}
\usepackage{ulem}
\usepackage{color}
\usepackage{dcolumn}
\usepackage{amsmath}
\usepackage{amssymb}
\usepackage{bm}
\usepackage{hyperref}
\usepackage{latexsym}
\usepackage{color}
\usepackage{subfigure}
\usepackage{breqn}
\usepackage{setspace}

\def\beq{\begin{equation}}
\def\eeq{\end{equation}}                         
\def\bea{\begin{eqnarray}}                           
\def\eea{\end{eqnarray}}

\begin{document}

\begin{center}
{\Large{\bf Domain Growth in the Active Model B: Critical and Off-critical Composition}} \\
\ \\
\ \\
by \\
Sudipta Pattanayak$^{1}$, Shradha Mishra$^2$ and Sanjay Puri$^3$ \\
$^1$S.N. Bose National Centre for Basic Sciences, JD Block, Sector III, Salt Lake City, Kolkata -- 700106, India. \\
$^2$Department of Physics, Indian Institute of Technology BHU, Varanasi -- 221005, India. \\
$^3$School of Physical Sciences, Jawaharlal Nehru University, New Delhi -- 110067, India. \\
\end{center}

\begin{abstract}
We study the ordering kinetics of an assembly of {\it active Brownian particles} (ABPs) on a two-dimensional substrate. We use a coarse-grained equation for the composition order parameter $\psi ({\bf r},t)$, where ${\bf r}$ and $t$ denote space and time, respectively. The model is similar to the {\it Cahn-Hilliard equation} or {\it Model B} (MB) for a conserved order parameter with an additional activity term of strength $\lambda$. This model has been introduced by Wittkowski et al., Nature Comm. {\bf 5}, 4351 (2014), and is termed {\it Active Model B} (AMB). We study domain growth kinetics and dynamical scaling of the correlation function for the AMB with critical and off-critical compositions. The quantity $P = \mbox{sign}(\lambda \times \psi_0)$ governs the asymptotic growth kinetics for the off-critical AMB, where $\psi_0$ denotes the average order parameter. For negative $P$, the domain growth law is the usual Lifshitz-Slyozov growth law with $L(t,\lambda) \sim t^{1/3}$. For positive $P$, the growth law shows a crossover to a novel growth law $L(t,\lambda) \sim t^{1/4}$. Further, the correlation function shows good dynamical scaling for the off-critical AMB but the scaling function has a dependency on $\psi_0$ and $\lambda$. We also study the effects of both additive and multiplicative noise on the AMB.
\end{abstract}

\newpage

\section{Introduction}
\label{introduction}

There has been intense recent interest in assemblies of self-propelled particles (SPPs), which arise in many natural systems \cite{ref1,ref2,ref3,ref4,ref5,ref6,ref7}. Each SPP converts energy into a systematic movement, thereby violating {\it time-reversal symmetry} \cite{ref8}. The scale of these assemblies ranges from micrometers \cite{Nedelec1997,Yokota1986,Bonner1998} to several meters \cite{Chen2019,Parrish1997,Helbing2000}. An important example of an SPP is an {\it active Brownian particle} (ABP), which is symmetric in shape but has a preferred direction of motion \cite{ref11,ref17,ref18}. Many lab-designed particles like active colloids \cite{ref3,ref4,ref5,ref6,ref7} and active Janus particles \cite{ref15,ref16} can be modeled as ABPs. These have many potential technological and pharmaceutical  applications, e.g., directional transport \cite{ref17,ref18,ref19}, sorting of particles \cite{ref20,ref21}, etc. In a recent study, Zottl et al. \cite{ref22stark} discussed generic features of the dynamics of active colloids in bulk and in confinement. These authors also reviewed the emergent collective behavior of active colloidal suspensions, focusing on their structural and dynamic properties.

One of the most interesting features of a collection of ABPs is that they show {\it motility-induced phase separation} (MIPS) \cite{ref11,ref12,ref13,ref14,thermo}. A recent study of Wittkowski et al. \cite{ref9} discusses the domain growth kinetics of an assembly of ABPs. These authors proposed a coarse-grained model analogous to the well-known {\it Cahn-Hilliard} (CH) equation or {\it Model B} (MB) for a conserved order parameter \cite{pw09}. This new model has been termed as {\it Active Model B} (AMB). In a recent study \cite{AMB}, we presented a detailed study of growth kinetics in the AMB with a critical composition.

In this paper, we revisit the problem of domain growth kinetics in the AMB. We present further results for the AMB with a critical composition. We discuss how the active term influences the scaling behavior of the correlation function for the critical mixture. Furthermore, we also present results for growth kinetics and dynamical scaling in the AMB with asymmetric or off-critical compositions. In this case, the growth kinetics depends on the sign of the product of the average order parameter ($\psi_0$) and activity strength ($\lambda$), $P = \mbox{sign}(\psi_0 \times \lambda)$. The size of the growing domains varies as $L(t,\lambda) \sim t^{1/z}$. For positive $P$, the growth exponent $z$ shows a crossover from the Lifshitz-Slyozov (LS) growth law ($z = 3$) at early times to an asymptotic value $z \simeq 4$ at late times. On the other hand, for negative $P$, the system behaves like MB and always obeys the LS growth law. Moreover, we note that the correlation function shows good dynamical scaling for the off-critical AMB. However, the scaling
function has a dependence on $\psi_0$ and $\lambda$. In this paper, we have also studied the effects of fluctuations on domain growth morphology in AMB. We consider cases with both additive and multiplicative noises for critical and off-critical compositions. Our results show that fluctuations do not significantly affect the domain morphology of the AMB. However, a detailed quantitative study is required to confirm the relevance of noise in the AMB.

This paper is organized as follows. In Sec.~\ref{Model}, we introduce the AMB and present details of our simulations and system parameters. In Sec.~\ref{results}, we discuss our numerical results. In Sec.~\ref{DMSC}, we present detailed results for the dynamical scaling of the correlation function in the critical AMB. The effects of additive and multiplicative noise on the critical AMB are discussed in Sec.~\ref{Noise}. In Sec.~\ref{DMDC}, the growth kinetics and scaling behavior for the off-critical AMB are discussed. The effects of fluctuations on the off-critical AMB are discussed in Sec.~\ref{NoiseOC}. Finally, Sec.~\ref{discussion} summarizes our main results.

\section{Model}
\label{Model}

We consider an assembly of ABPs on a two-dimensional $(d = 2)$ substrate and study the dynamics of the system. The ABPs move with a self-propulsion speed $v_0$, and their characteristic rotation frequency is $\tau^{-1}$. The local density of the ABPs is denoted as $n(r, t)$, where $r$ is the position vector. The corresponding order parameter is $\psi({\bf r}, t) = 2n({\bf r}, t)-1,$ so that the regions with $\psi > 0$ are enriched in particles. Since the density of particles is conserved, the dynamical equation for the local order parameter is expressed as: 
\begin{equation}
	\frac{\partial \psi({\bf r},t)}{\partial t}=-\nabla \cdot {\bf J} ,
	\label{eqn1}
\end{equation}
\begin{equation}
	{\bf J}=-\nabla \mu ,
	\label{eqn2}
\end{equation}
\begin{equation}
	\mu = -\psi({\bf r},t) + \psi({\bf r},t)^{3} - \nabla^{2}\psi({\bf r},t) + \lambda|\nabla \psi({\bf r},t)|^{2} . 
	\label{eqn3}
\end{equation}

The above equations are formulated in dimensionless units which are obtained by rescaling length and time by the persistence length $v_{0} \tau$ and the relaxation time $\tau$, respectively, for all the particles.   
Eq.~(\ref{eqn1}) expresses the conservation of density of particles.
Eq.~(\ref{eqn2}) states that the mean current of order parameter is proportional to
a nonequilibrium  chemical potential $\mu$,
as shown in Eq.~(\ref{eqn3}). 
$\mu$ is the sum of bulk $\mu_{0}$ and gradient $\mu_{1}$ contributions.
The bulk part is the same as that in Model B, $\mu_{0} = -\psi({r},t) + \psi({\bf r},t)^{3}$,
which is derived from the bulk free-energy density 
\begin{equation}
F_{0}=-\frac{\psi({\bf r},t)^{2}}{2} + \frac{\psi({\bf r},t)^{4}}{4} 
\end{equation}
of a symmetric $\psi^{4}$ field theory.
The gradient term can be written as the sum of two terms, 
$\mu_{1} = \mu_{1}^{p} + \mu_{1}^{a}$,
where $\mu_{1}^{p}$ is passive integrable term and $\mu_{1}^{a}$ is the active term. 
We choose $\mu_{1}^{p}$ same as standard Model B (MB), 
$\mu_{1}^{p}=-\nabla^{2}\psi({\bf r},t)$ and its corresponding gradient term in the Ginzburg-Landau (GL) 
free energy density is 
\begin{equation}
F_{1}=\frac{\vert(\nabla \psi({\bf r},t))\vert^{2}}{2} .
\end{equation}
In the non-integrable part, $\mu_{1}^{a}=\lambda|\nabla \psi({\bf r},t)|^{2}$, $\lambda$ represents \textit{activity} of the ABPs and
it is also a tunable parameter in our study.
The origin of  non-integrable $\lambda-$term in 
 the model is similar to the lowest-order nonlinear interfacial  diffusion
 term in KPZ equation\cite{kardar}.
Therefore, we express the update equation for the order parameter 
$\psi({\bf r}, t)$ for the Active Model B (AMB) as,
\begin{equation}
\frac{\partial \psi({\bf r},t)}{\partial t} = \boldsymbol{\nabla} \cdot \left\lbrace \boldsymbol{\nabla}[-\psi({\bf r},t) + \psi({\bf r},t)^{3} - \nabla^{2}\psi({\bf r},t) + \lambda | \boldsymbol{\nabla} \psi ({\bf r}, t)|^{2}] \right\rbrace .
\label{eqn_activemodelB}
\end{equation}

We should stress that microscopic models of active matter are anisotropic on short time and length scales. However, the coarse-graining procedure used to obtain the AMB smears out this anisotropy and the resultant partial differential equation model is isotropic in nature. We numerically integrate Eq.~(\ref{eqn_activemodelB}) on a two-dimensional substrate of size $N^2$ with periodic boundary conditions (PBCs). For mean value of order parameter, $\psi_0=0$ (symmetric composition: equal number of occupied and empty sites), the model is termed critical AMB whereas we refer to it as the off-critical AMB for $\psi_0 \ne 0.0$ (asymmetric composition). In numerical simulations, $\lambda$ is varied from $0.0$ to $4.0$. 
However, for $\lambda=0.0$, the model reduces 
to the standard MB \cite{pw09}. We solve Eq.~(\ref{eqn_activemodelB}) via a simple Euler discretization scheme with mesh sizes $\Delta t = 0.01$ and $\Delta x=1.0$. We consider system sizes $N = 512 \times 512$ and all observables are averaged over 100 independent runs. 

Furthermore, we analyze the effects of thermal fluctuations in the AMB by introducing noise (either additive or multiplicative) in the coarse-grained Eq.~(\ref{eqn_activemodelB}). The update Eq.~(\ref{eqn_activemodelB}) is modified in the presence of noise as
\begin{dmath}
	\frac{\partial \psi({\bf r},t)}{\partial t} = \boldsymbol{\nabla} \cdot \left\lbrace \boldsymbol{\nabla}[-\psi({\bf r},t) + \psi({\bf r},t)^{3} - \nabla^{2}\psi({\bf r},t) - \lambda |\boldsymbol{\nabla} \psi({\bf r}, t)|^{2}] + \eta{\bf f}({\bf r},t) \right\rbrace,
\label{eqn_add}
\end{dmath}
where the strength of the noise is $\eta= \eta_a$ for {\it additive noise} (AN). We also consider the case of {\it multiplicative noise} (MN), where
\begin{equation}
\eta = \eta_a \left(\frac{1+\psi({\bf r},t)}{2}\right)^{1/2} .
\end{equation}
In Eq.~(\ref{eqn_add}), ${\bf f}({\bf r},t)$ denotes Gaussian white noise with mean zero and unit variance.
The form of the multiplicative noise is similar to that introduced by Dean for interacting Brownian particles \cite{dean}.

\section{Numerical Results}
\label{results}

\subsection{Critical Mixtures}
\label{critical}

\subsubsection{Domain Morphology and Dynamical Scaling}
\label{DMSC}

\begin{figure}[htb]
\centering
\includegraphics[width=0.7\linewidth]{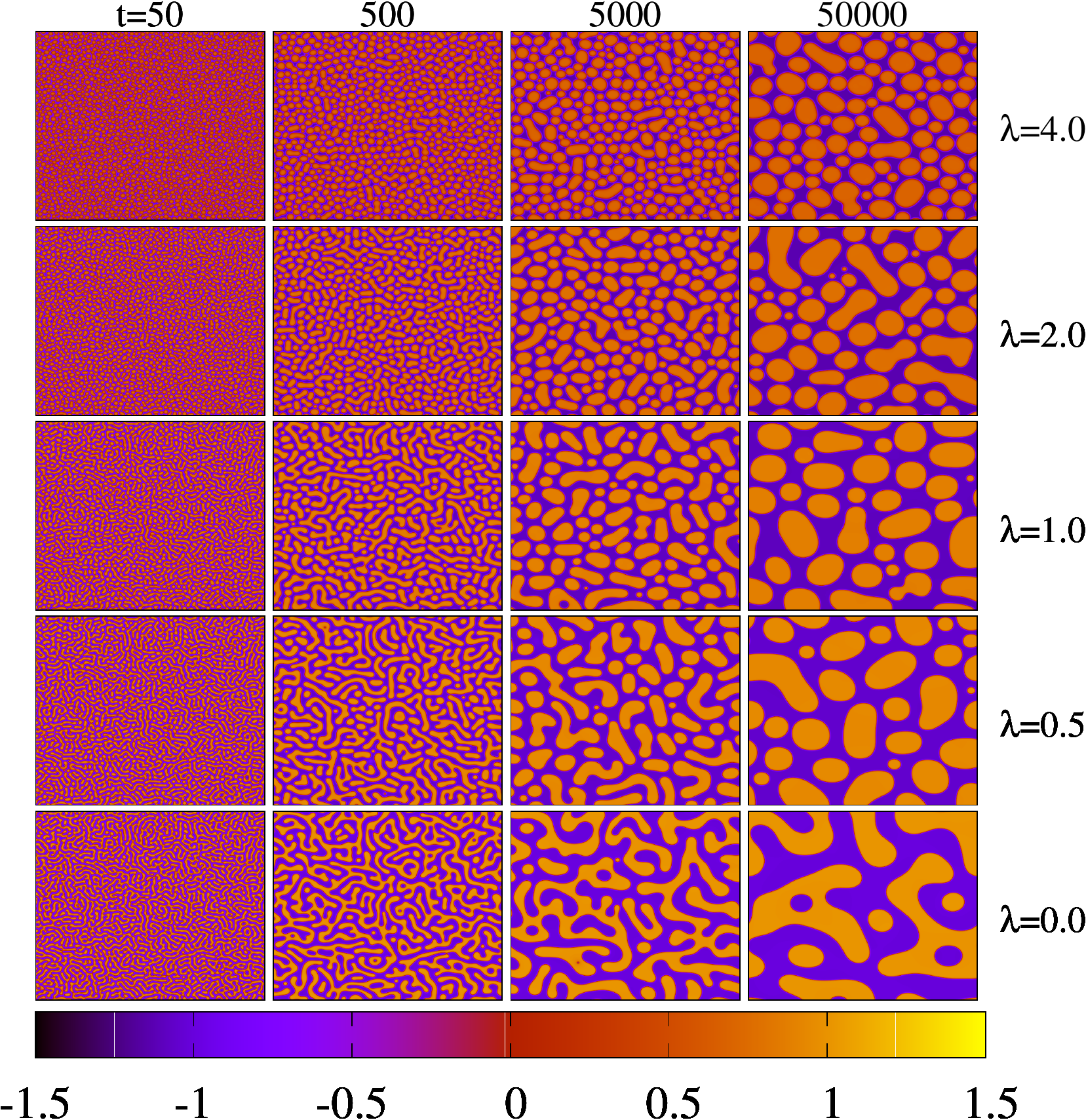}
\caption{(Color online) Evolution snapshots of the system at different times for various values of activity $(\lambda)$ as shown in Fig.~1 in our recent study \cite{AMB}. The system size is $N^2 = 512^2$, and the mean order parameter is $\psi_0=0.0$, corresponding to a critical composition. The color bar represents the range of $\psi$-values.}
\label{fig_1}
\end{figure}

First, we study the domain morphology and dynamical scaling of the correlation function and structure factor for the critical AMB, $\psi_0=0.0$, for different activity $\lambda$. Snapshots of the local order parameter $\psi({\bf r}, t)$ with time for different activity $(\lambda=0.0,0.5,1.0$, $2.0$ and $4.0)$ are shown in Fig.~\ref{fig_1}. The empty and filled regions denote $\psi < 0$ and $\psi > 0$, respectively. For all activity $\lambda$, we initially start with a homogeneous mixed state and the domains of $\psi$-rich regions start to grow with time. We find that the domains are bi-continuous for $\lambda=0$ and these domains become isolated in the presence of activity $\lambda$. We also note that the system is symmetric under the transformation $\psi = -\psi$, and $\lambda = -\lambda$.

To further quantify the domain morphology, we calculate the order parameter correlation function 
\begin{equation}
C(r,t)=\langle \psi({\bf r}_0 + {\bf r}, t) \psi({\bf r}_0, t) \rangle .
\end{equation}
We also compute the corresponding structure factor
\begin{equation}
S(k, t) = \langle |\psi({\bf k},t)|^2 \rangle ,
\end{equation}
where $\psi({\bf k}, t)$ is the Fourier transform of the local order parameter $\psi({\bf r}, t)$ at wave-vector ${\bf k}$. The angular brackets $\langle .. \rangle$ denote an average over the reference positions ${\bf r}_0$ and 100 independent realizations, followed by a spherical averaging. We find that the size of the domains increases with time. The growth of the domains
is measured by calculating the domain scale $L(t,\lambda)$ which is defined as a characteristic length where the correlation function $C(r,t)$ decays to $0.5$.

In our recent study \cite{AMB}, we found that the ordering kinetics of growing domain shows a crossover from early-time LS growth $L(t,\lambda) \sim t^{1/3}$ to a late-time novel growth law $L(t,\lambda) \sim t^{1/4}$ \cite{pbl97,gbp05}. Hence, the general expression for $C(r,t)$ will be a function of activity and time both. Furthermore, we consider the following form for the correlation function $C(r, t, \lambda)$ in the presence of activity,
\begin{equation}
	C(r, t, \lambda) = C_0 C\left(\frac{r}{L(t,\lambda)}\right), 
\end{equation}
where $C_0$ is a constant pre-factor and  $L(t,\lambda)$ is the characteristic length. As proposed in \cite{AMB}, for a finite $\lambda$, we can write $L(t,\lambda) = t^{1/3}f(t/t_{c})$, where $f(x)$ is the scaling function and $t_{c}$ is the crossover time which varies as $\lambda^{-3/2}$. Since for $t<t_c$,  $L(t,\lambda) \sim t^{1/3}$, hence, $f(x)$ is constant for $x \rightarrow 0$. As the asymptotic growth dynamics is $L(t,\lambda) \sim t^{1/4}$, hence, $f(x) = x^{-1/12}$ for $x \rightarrow \infty$ \cite{AMB}.

\begin{figure}[htb]
\centering
\includegraphics[width=0.5\linewidth]{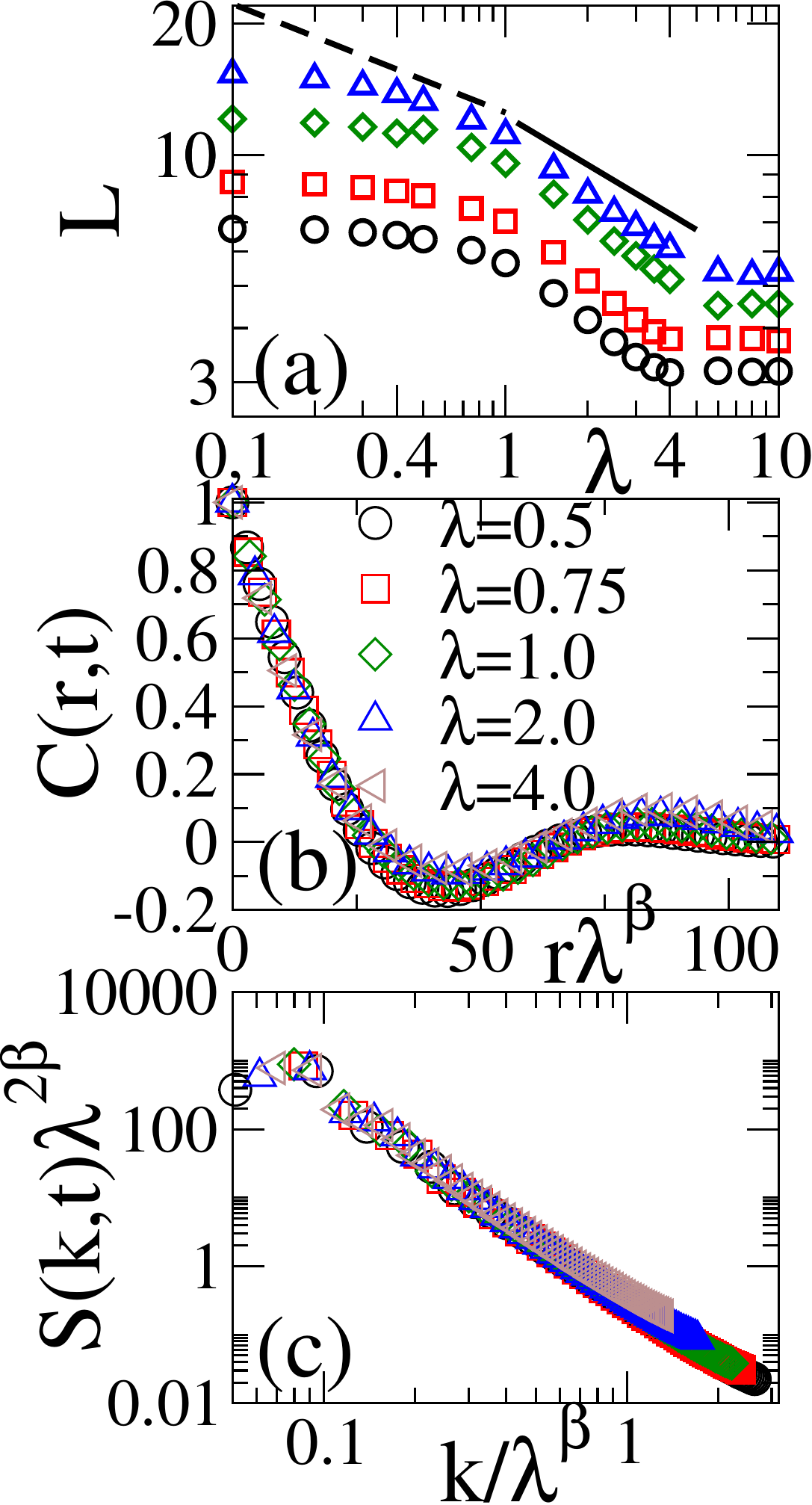}
\caption{(Color online) (a) Log-log plot of $L(t,\lambda)$ vs. $\lambda$ at different times. Circles, squares,
diamonds and triangles denote $t=2500,5000,12500$ and 25000, respectively. Dashed and solid lines represent slope $1/4$ and $3/8$, respectively. (b)-(c) Static scaling plots of $C(r,t)$ vs. $r \lambda^{\beta}$ and $S(k,t)\lambda^{2\beta}$ vs. $k/\lambda^{\beta}$ for different values of $\lambda$ at $t=25000$, respectively.
$\beta = 1/4$ for $\lambda = 0.5$ and $0.75$ and it is 3/8 for $\lambda = 2.0$ and $4.0$. The average order parameter  $\psi_0 = 0$, corresponding to a critical composition. The system size is $N^2=512^2$.}
\label{fig_2}
\end{figure}

Moreover, using simple scaling arguments, we also estimate the variation of the characteristic length $L(t,\lambda)$ with $\lambda$ at a fixed time. For high $\lambda$,  we find that the crossover happens at early time, hence, $L(t,\lambda) \sim t^{1/4}$. Therefore, at a fixed time, the characteristic length $L(t,\lambda)$ varies as $L(t,\lambda) \sim \lambda^{-3/8}$. However, for moderate $\lambda (<1$), the crossover happens over a long period of time ($\sim 1000$), hence, it is difficult to extract the simple scaling relation between $L(t,\lambda)$ and the activity. However, we extract the relation between $L(t,\lambda)$ and $\lambda$ numerically and find two distinct scaling relation for moderate and high $\lambda$. In Fig.~\ref{fig_2}(a), we show the variation of $L(t,\lambda)$ vs. $\lambda$ at four different times.  $L(t,\lambda) \sim \lambda^{-1/4}$ for moderate $\lambda~(\lambda < 1)$ and $L(t,\lambda) \sim \lambda^{-3/8}$ for high $\lambda~(\lambda > 1)$. The system does not exhibit coexisting domains for $\lambda > 4$ \cite{ref9}. Thus, $L(t,\lambda)$ is flat in that regime, as shown in Fig.~\ref{fig_2}(a). In Fig.~\ref{fig_2} (b), (c), we show the scaling for the correlation function [$C(r,t)$ vs. reduced distance $r \lambda^{\beta}$] and structure factor [$S(k,t) \lambda^{2 \beta}$ vs. reduced wave-vector $k/\lambda^{\beta}$], respectively. We find good scaling collapse of data for moderate and high activity with $\beta$ $\sim 1/4$ and $\sim 3/8$, respectively.

\subsubsection{Effects of Noise}
\label{Noise}

In the previous section, we studied the domain morphology and scaling behavior for the critical AMB without any fluctuations. To study the effects of fluctuations on domain growth morphology, we consider additive and multiplicative noise, as introduced in Eq.~(\ref{eqn_add}). In general, noise in the coarse-grained equation is multiplicative in nature. The leading order expression for the multiplicative noise is obtained by the derivation of the coarse-grained equation for the local order parameter starting from the microscopic Langevin equation, as shown in \cite{dean} for the interacting Brownian particle system. Hence, the leading order contribution of  noise in our model depends on $\psi({\bf r}, t)$, as given in Eq.~(\ref{eqn_add}).
It is higher in magnitude where the local order parameter is higher and vice versa. Since the density fluctuations in active systems are large \cite{ref11, GNF1, ref30, ref28, GNF5}, the multiplicative noise will act inhomogeneously in the system.

\begin{figure}[htb]
\centering
\includegraphics[width=0.6\linewidth]{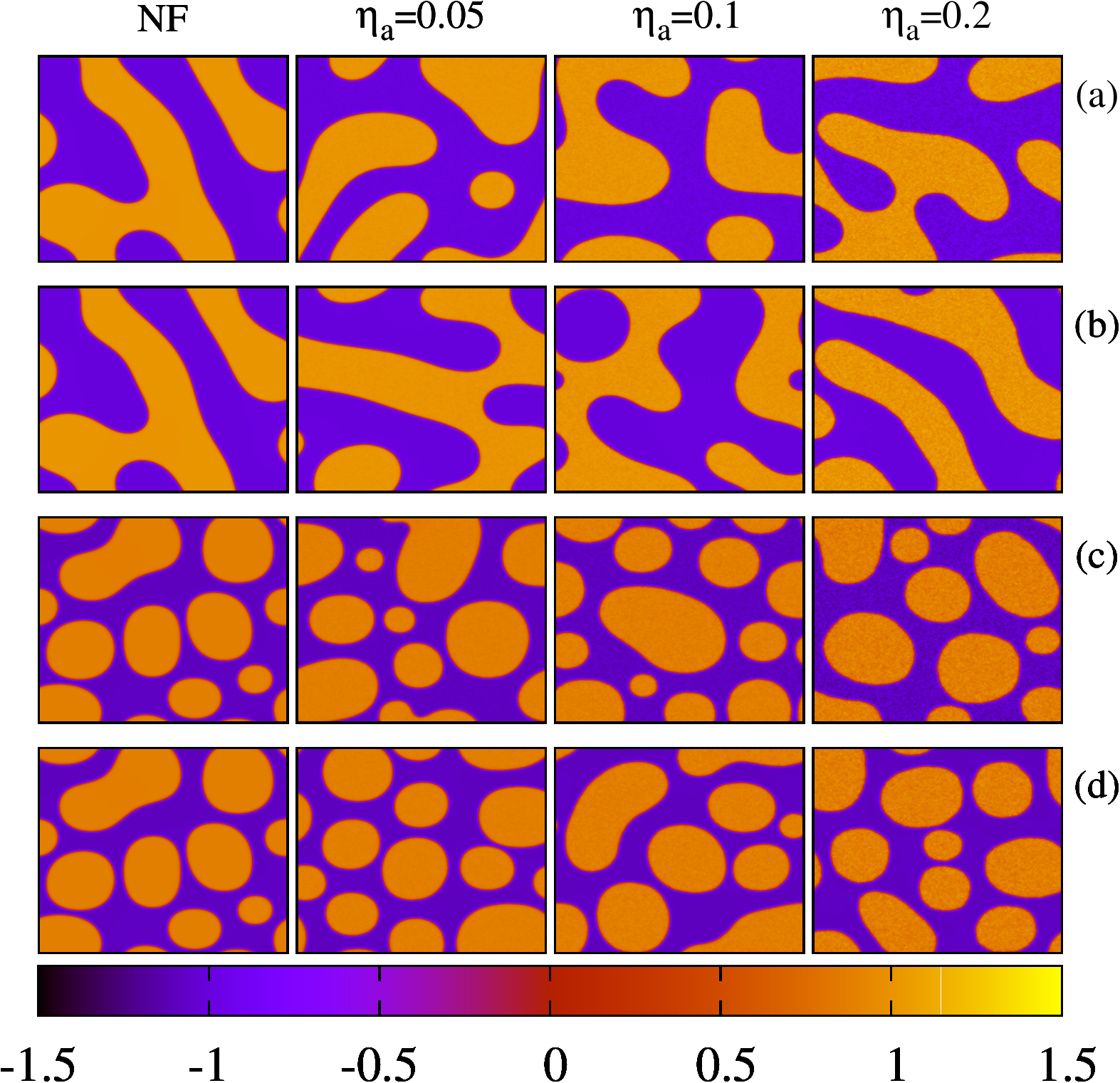}
	\caption{(Color online) (a) and (b), Evolution snapshots of the system for the MB 
        (corresponding to a composition $\psi_0=0$ with activity $\lambda=0$)
        in the presence of additive noise and multiplicative noise, respectively, 
        at $t=50000$. (c) and (d), Evolution snapshots of the system for the active model
        B (corresponding to a composition $\psi_0=0$ with activity $\lambda=1$)
        in the presence of additive noise and multiplicative noise, respectively,
        at $t=50000$. Four columns represent noise strength $\eta_a=0.0$ or noise free 
        system (NF), $0.05, 0.1$ and $0.2$, respectively. 
        The system size is $N^2= 512^2$. The color bar denotes the range of $\psi$-values.}
\label{fig_3}
\end{figure}

We consider the effects of both additive and multiplicative noise on domain growth in the critical MB and the critical AMB. We plot the evolution snapshots at $t=50000$ without noise or $\eta_{a}=0$ (NF), and for non-zero values of $\eta_{a}$ in Fig.~\ref{fig_3}. In panel (a) and (b) of Fig.~\ref{fig_3}, the evolution snapshots of the system in the presence of additive (AN) and multiplicative (MN) noises for the critical MB are shown respectively. Similarly, the evolution snapshots of the system in the presence of additive (AN) and multiplicative (MN) noises for the critical AMB are shown in panel (c) and 
(d) of Fig.~\ref{fig_3}, respectively. Surprisingly, there are no significant differences in the domain structures in the presence of noise. Therefore, the fluctuations do not affect the domain morphology of the MB or the AMB. This has been known earlier in the context of MB, where it is customary to neglect thermal fluctuations as these are irrelevant in the asymptotic regime \cite{po88}. However, for the AMB, this requires a more quantitative characterization which we leave to future work. It is worth mentioning that fluctuations are known to affect the growth kinetics in microscopic active models \cite{Chate}.

\subsection{Off-critical Mixtures}
\label{off-critical}

\subsubsection{Domain Morphology, Growth Kinetics and Dynamical Scaling}
\label{DMDC}

\begin{figure}[htb]
\centering
\includegraphics[width=0.6\linewidth]{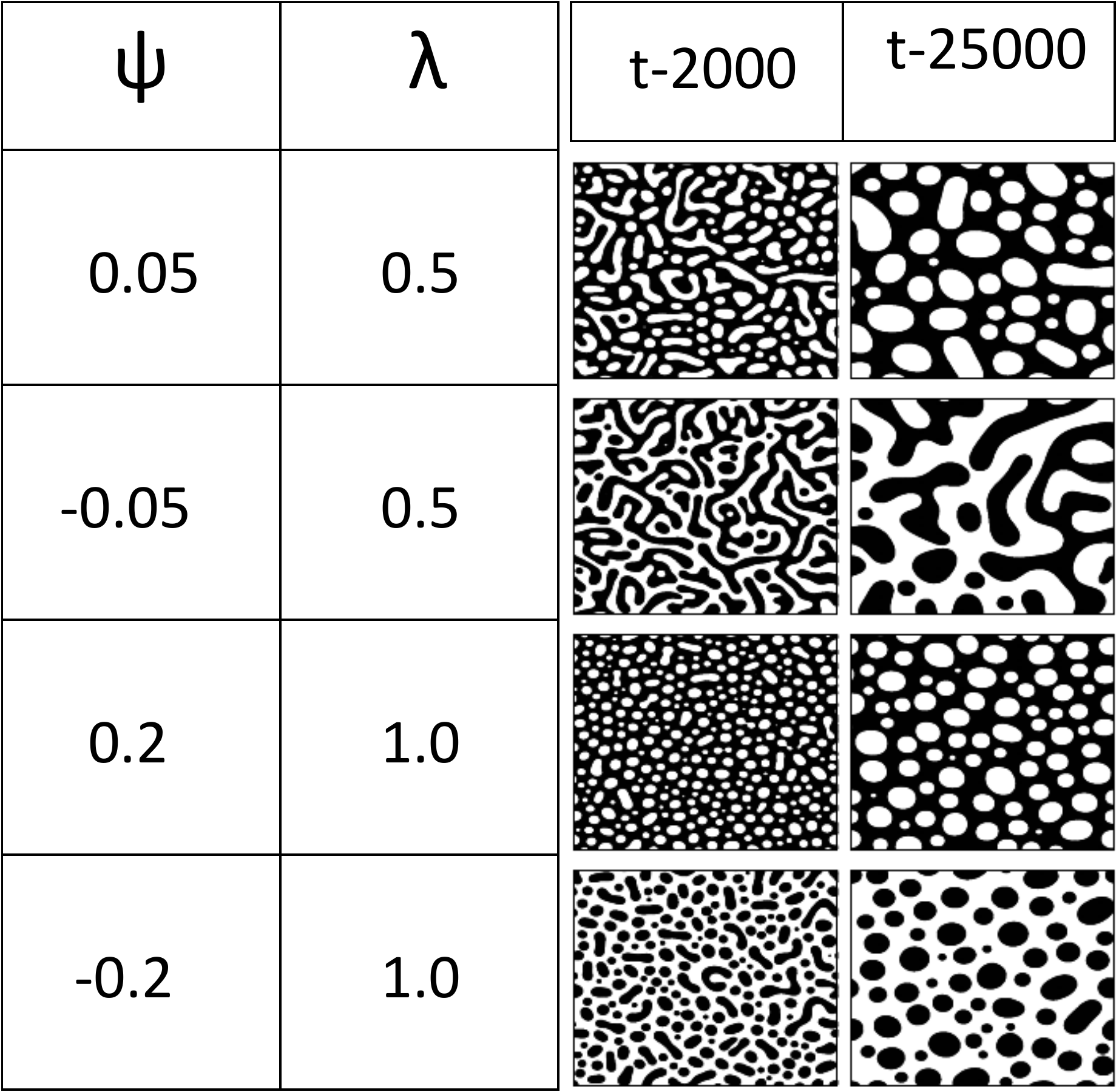}
\caption{Evolution snapshots of the system at $t=2000$ and $25000$, corresponding to off-critical composition
	for $\psi_0=0.05$ and $\psi_0=0.2$ . $\lambda=0.5$ and $1.0$ are considered for
        $\psi_0=0.05$ and $\psi_0=0.2$, respectively.
        Empty and filled regions represent $\psi < \psi_0$ and $\psi > \psi_0$ , respectively.
	The system size is $N^2=512^2$.}
\label{fig_4}
\end{figure}

In this section, we study the effects of activity on the ordering kinetics of   
 the off-critical AMB, $\psi_0 \neq 0.0$. We characterize the effects for both (i) slightly off-critical 
$\psi_0 = \pm0.05$ and (ii) highly off-critical $\psi_0 =\pm 0.2$.
In Fig.~\ref{fig_4} , we show the snapshots of the system at early time $t=2000$ and at late time $t=25000$  for different combinations of $\psi_0$ and activity parameter $\lambda$. For the off-critical AMB, we note that the system is no longer symmetric under the $\psi = -\psi$ and $\lambda = -\lambda$ transformation. The dynamics of the off-critical AMB depends on the relative sign of $\psi_0$ and $\lambda$. Therefore, we define the relative sign function $P=\mbox{sign}(\psi_0 \times \lambda)$.

\begin{figure}[htb]
\centering
\includegraphics[width=0.6\linewidth]{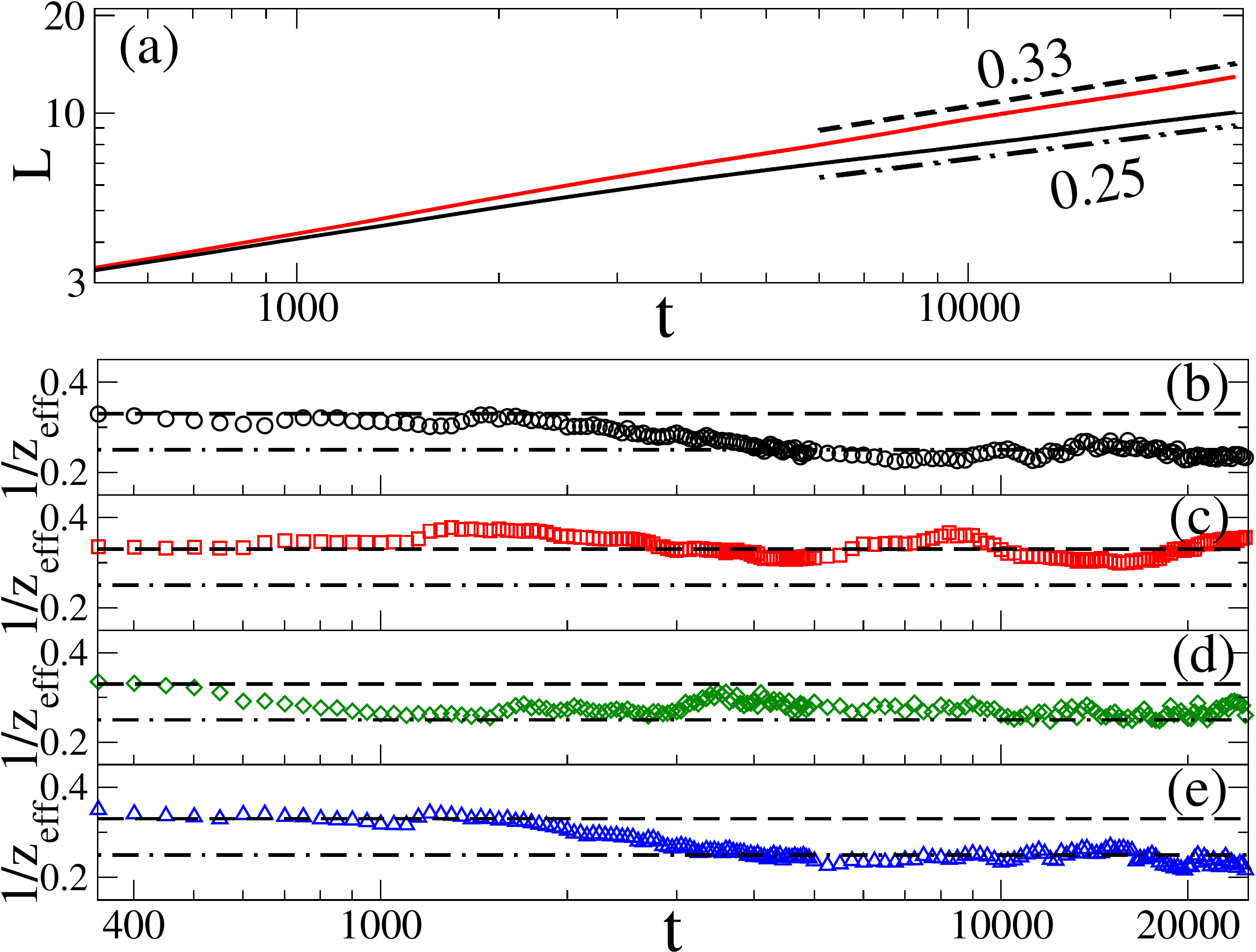}
        \caption{(Color online) (a) Log-log plot of $L(t,\lambda)$ vs. $t$. Black and red lines
        represent $\psi_0 = 0.05$ and $-0.05$, respectively. Activity $\lambda = 0.5$.
	The dashed and dot-dashed lines denote the growth laws $L(t,\lambda) \sim t^{1/3}$ and
	$L(t,\lambda) \sim t^{1/4}$, respectively.
	(b)-(e) Plot of effective exponent ($1/z_{\rm eff}$) vs. $t$ for different $\lambda$ and $\psi_0$.
        Circles, squares, diamonds and triangles represent
	$(\psi_0, \lambda)  = (0.05, 0.5), (-0.05, 0.5), (0.2, 1.0)$ and $(-0.2, 1.0)$ respectively.
	In each frame, the horizontal dashed line denotes $1/z_{\rm eff} = 1/3$, and the dot-dashed 
	line denotes $1/z_{\rm eff} = 1/4$. The system size is $N^2 = 512^2$.}
\label{fig_5}
\end{figure}

For the slightly off-critical mixture, the domains are 
bi-continuous when $\psi_0$ and $\lambda$ appear with the opposite signs (negative $P$), whereas these domains 
becomes droplet-like for the same signs of $\psi_0$
and $\lambda$ (positive $P$), as shown in the top two panels in Fig.~\ref{fig_4}. 
However, the domains are
always isolated irrespective of the signs of $\psi_0$ and $\lambda$ 
for the high off-critical mixture, as shown in the bottom two panels in Fig.~\ref{fig_4}.
For the slightly off-critical mixture ($\psi_0=0.05$), we also show the variation
of $L(t,\lambda)$ with time $t$ in Fig.~\ref{fig_5}(a), for the same and opposite signs of $\psi_0$ and
$\lambda$. We note that $L(t,\lambda) \sim t^{1/z}$ and the asymptotic growth exponent $z$ is $4$ and $3$ for the same and opposite signs of
$\psi_0$ and $\lambda$, respectively. 
Moreover, we estimate $1/z_{\rm eff}$ from numerical calculation to understand the domain growth
for different combinations of $\psi_0$ and $\lambda$, which is defined
as,
\begin{equation}
	\dfrac{1}{z_{\rm eff}}=\dfrac{d \ln L(t,\lambda)}{d \ln t}.
        \label{eqnexpo}
\end{equation}
For the same sign of $\psi_0$ and $\lambda$, $1/z_{\rm eff}$ shows a crossover from
an early time value $1/3$ to an asymptotic value $1/4$ for the slightly off-critical mixture, 
as shown in Fig.~\ref{fig_5}(b). 
However, $1/z_{\rm eff}$ always remains close to $1/3$ if $\psi_0$
and $\lambda$ have the opposite signs for the slightly off-critical mixture, as shown in Fig.~\ref{fig_5}(c). 
It suggests that the
system behaves more like the critical MB due to the competition between 
the off-criticality and the activity.
However, the highly off-critical mixture shows the switching from an early time $1/3$
growth to an asymptotic $1/4$ growth
behavior similar to the critical AMB \cite{AMB}, as shown in Fig.~\ref{fig_5}(d) and (e).

\begin{figure}[htb]
\centering
\includegraphics[width=0.5\linewidth]{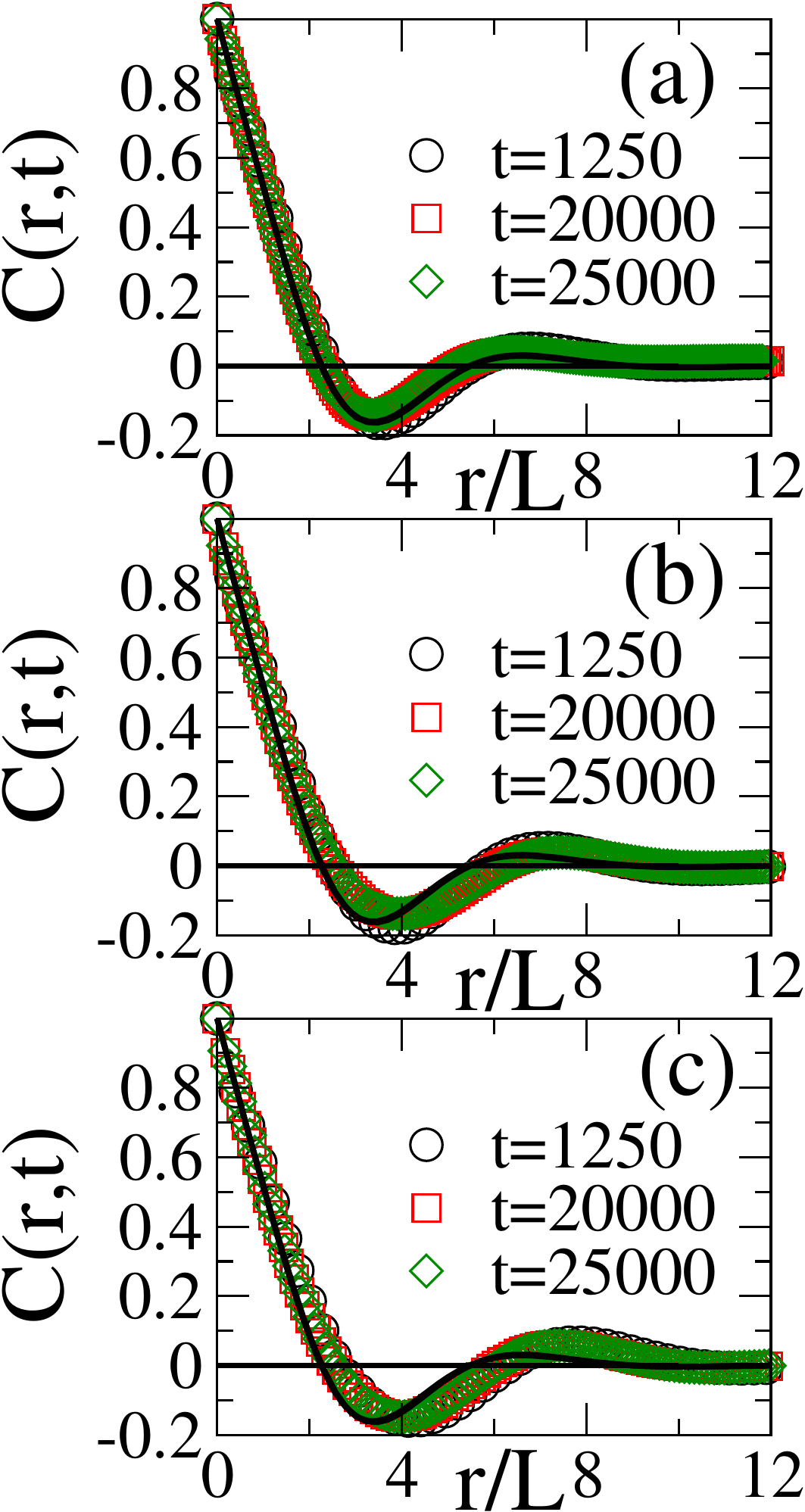}
	\caption{(Color Online) (a) Scaling plot of $C(r,t)$ vs. $r/L(t,\lambda)$ corresponding to 
        off-critical composition for $\psi_0 = -0.05$ and the activity 
        $\lambda = 0.5$. We plot data at the indicated times. 
        The solid line denotes the scaling function for the Model B 
        at critical composition $\psi_0 = 0$. (b) Analogous to (a) but 
        $\psi_0 = 0.05$. (c) Analogous to (a) but $\psi_0 = 0.2$ 
        and $\lambda = 1.0$. The system size is $N^2= 512^2$.}
\label{fig_6}
\end{figure}

Furthermore, we study dynamical scaling for the off-critical cases. In
the previous section, we have mentioned that the growth kinetics of the off-critical AMB
depends on the relative signs of $\psi_0$ and $\lambda$. In Fig.~\ref{fig_6} (a)-(c),
the dynamical scaling of the correlation function for the slightly off-critical
composition with $\psi_0=-0.05$ and $\lambda=0.5$; $\psi_0=0.05$ and $\lambda=0.5$;
and highly off-critical composition with $\psi_0=0.2$ and $\lambda=1.0$ are shown
respectively. We have also compared the dynamical scaling of the off-critical AMB
with the critical MB which is shown by black lines in the Fig.~\ref{fig_6} (a)-(c).
We note that the off-critical AMB always show a good dynamical scaling for both 
slightly off-critical and highly off-critical compositions with all
combination of signs of $\psi_0$ and $\lambda$. However, interestingly, we find that 
the dynamical scaling of the slightly off-critical AMB behaves more like
the critical MB due to the competition between the off-criticality and the activity, 
as shown in Fig.~\ref{fig_6} (a).
However, the dynamical scaling for the slightly off-critical AMB with same signs 
of $\psi_0$ and $\lambda$ and for highly off-critical compositions deviates from the 
critical MB as shown in Fig.~\ref{fig_6}(b) and (c). Also, the deviation is 
proportional to the strength of the off-criticality and activity. We observe similar behavior for the critical AMB in our recent study \cite{AMB}.

\subsubsection{Effects of Noise}
\label{NoiseOC}

\begin{figure}[htb]
\centering
\includegraphics[width=0.6\linewidth]{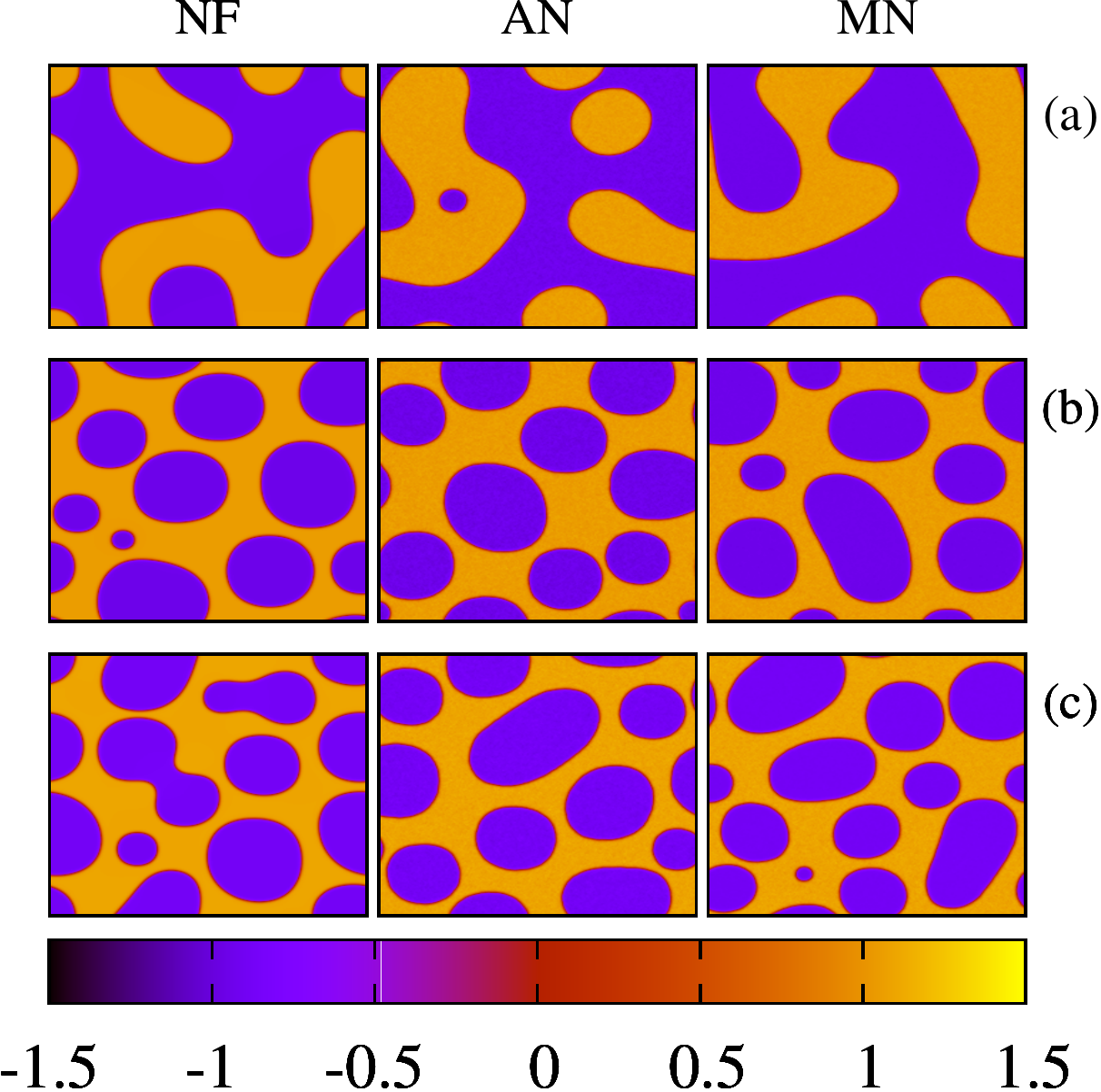}
	\caption{(Color online) Evolution snapshots of the system for 
	the off-critical AMB (a) corresponding to a composition $\psi_0=-0.05$ 
        with activity $\lambda=0.5$, (b) corresponding to a composition 
        $\psi_0=0.05$ with activity $\lambda=0.5$, and (c) corresponding to a 
        composition $\psi_0=0.2$ with activity $\lambda=1.0$  
	at $t=50000$. Three columns represent noise free system (NF), 
        system in the presence of 
	additive noise (AN) of strength $\eta_a=0.1$ 
        and multiplicative noise (MN) of strength
        $\eta_a=0.1$, respectively. The system size is $N^2= 512^2$. 
        The color bar denotes the range of $\psi$-values.}
\label{fig_7}
\end{figure}

In Sec.~\ref{Noise}, we studied the effects of fluctuations
on domain growth kinetics and morphology for the critical AMB.
In this section, we study the effects
of additive and multiplicative noise, as introduced in Eq.~(\ref{eqn_add}),
in the off-critical AMB. 
We plot the steady-state snapshots of $\psi({\bf r}, t)$
without noise (NF), in the presence of additive (AN) and 
multiplicative (MN) noises for the slightly off-critical AMB with 
opposite signs of $\psi_0$ and $\lambda$ ($\psi_0=-0.05$, $\lambda=0.5$);
the slightly off-critical AMB with 
same signs of $\psi_0$ and $\lambda$ ($\psi_0=0.05$, $\lambda=0.5$);
and highly off-critical AMB with $\psi_0=0.2$, and $\lambda=1.0$
in Fig.~\ref{fig_7} (a), (b) and (c), respectively.

As in the critical AMB, there are no significant differences in the domain structures in the presence of noise. 
Therefore, the fluctuations do not significantly affect the domain morphology of the off-critical AMB. However, as in the case of critical mixtures, a conclusive statement in this regard can only be made through a quantitative study of correlation functions, structure factors, growth laws, and other characteristic quantities.

\section{Summary and Discussion}
\label{discussion}

We have performed a detailed study on the growth kinetics of the Active Model B for both critical and off-critical mixtures. We summarize our main results as follows: we first studied domain morphology, and scaling behavior of the correlation function and the structure factor for the critical composition for different strengths of the activity $\lambda$. For the critical composition, the system is symmetric with respect to the sign of the activity term. For zero activity or the MB, the system shows the formation of bi-continuous domains. However, these domains are no longer bi-continuous as we introduce the activity in the system and the domains slowly become isolated. Furthermore, we note that
the correlation function and structure factor show two distinct scaling forms
for the moderate and high activity of the particles. Therefore,
the growth and morphology of the domains depends on the activity parameter.
Moreover, we study the effects of fluctuations on domain morphology
by introducing additive and multiplicative noises in the critical AMB.
Surprisingly, we note that the fluctuations do not affect the domain morphology 
of the MB and as well as of the critical AMB.

We also discussed the growth dynamics and dynamical scaling
of the correlation function for the off-critical composition or the off-critical AMB. 
For the slightly off-critical composition, the growth dynamics of the system depends
on the relative sign of $\psi_0$ and $\lambda$. The domains are isolated when $\psi_0$ and
$\lambda$ have the same signs, whereas the domains become bi-continuous, similar to the MB, for the
opposite signs of $\psi_0$ and $\lambda$. Therefore, for the slightly off-critical mixture,
there exists a competition between the asymmetry due to $\psi_0$ and the activity $\lambda$. 
However, the domains are always isolated for the high off-critical composition of the AMB.
Furthermore, we note that the correlation function shows 
good dynamical scaling for the off-critical AMB, similar to the critical AMB as shown in 
\cite{AMB}. However, the scaling function has a dependency on $\psi_0$ and $\lambda$.
Similar to the critical AMB, we have analyzed the effects of fluctuations using 
additive and multiplicative noises. We note that the fluctuations do not affect the domain
morphology of the off-critical AMB.

Our study opens a new direction to study the effects of activity on the domain growth in other active systems.
It will also be interesting to characterize the effects of noise quantitatively for these systems. We would also like to examine the possibility of rough interfaces due to the presence of quenched disorder \cite{ref43,ref44} in both theoretical and experimental studies.

\subsubsection*{Acknowledgments}

We are grateful to the TUE computational facility at S.N. Bose National Centre for Basic Sciences, Kolkata. S. Mishra thanks SERB (India) for financial support via project ECR/2017/000659. S. Pattanayak and S. Mishra would like to thank the Department of Physics, Indian Institute of Technology (BHU), Varanasi and S.N. Bose National Centre for Basic Sciences, Kolkata for their kind hospitality.

\newpage

\begin{thebibliography}{99}
	
\bibitem{ref1} M.C. Marchetti, J.F. Joanny, S. Ramaswamy, T.B. Liverpool, J. Prost, M. Rao and R. Aditi Simha, Rev. Mod. Phys. {\bf 85}, 1143 (2013).

\bibitem{ref2} M.E. Cates, Rep. Prog. Phys. {\bf 75}, 042601 (2012).

\bibitem{ref3} J.R. Howse, R.A.L. Jones, A.J. Ryan, T. Gough, R. Vafabakhsh and R. Golestanian, Phys. Rev. Lett. {\bf 99}, 048102 (2007).

\bibitem{ref4} S.J. Ebbens and J.R. Howse, Soft Matter {\bf 6}, 726 (2010).

\bibitem{ref5} S. Thutupalli, R. Seemann and S. Herminghaus, New J. Phys. {\bf 13}, 073021 (2011).

\bibitem{ref6} G. Volpe, I. Buttinoni, D. Vogt, H. Kummerer and C. Bechinger, Soft Matter {\bf 7}, 8810 (2011).

\bibitem{ref7} J. Palacci, S. Sacanna, A.P. Steinberg, D.J. Pine and  P.M. Chaikin, Science {\bf 339}, 936 (2013).

\bibitem{ref8} J. Jeong, S. Yvan, A. Louat, V. Brouet and P. Bourges, Nature Comm. {\bf 8}, 15119 (2017).

\bibitem{Nedelec1997} F. Nedelec, T. Surrey, A. C. Maggs and S. Leibler, Nature (London) {\bf 389}, 305 (1997).

\bibitem{Yokota1986} H. Yokota, private communication; Y. Harada, A. Noguchi, A. Kishino and T. Yanagida, Nature (London) {\bf 326}, 805 (1987); Y. Toyoshima et al., Nature (London) {\bf 328}, 536 (1987); S.J. Kron and J.A. Spudich, Proc. Natl. Acad. Sci. U.S.A. {\bf 83}, 6272 (1986).

\bibitem{Bonner1998} J.T. Bonner, Proc. Natl. Acad. Sci. U.S.A. {\bf 95}, 9355 (1998); M.T. Laub and W.F. Loomis, Mol. Biol. Cell {\bf 9}, 3521 (1998).

\bibitem{Chen2019} D. Chen, Y. Wang, G. Wu, M. Kang, Y. Sun and W. Yu, Chaos {\bf 29}, 113118 (2019).

\bibitem{Helbing2000}  D. Helbing, I. Farkas and T. Vicsek, Nature (London) {\bf 407}, 487 (2000); Phys. Rev. Lett. {\bf 84}, 1240 (2000).

\bibitem{Parrish1997} J.K. Parrish and W.M. Hamner, {\it Three Dimensional Animal Groups}, Cambridge University Press, Cambridge (1997).

\bibitem{ref11} Y. Fily and M.C. Marchetti, Phys. Rev. Lett. {\bf 108}, 235702 (2012).

\bibitem{ref17} S. Pattanayak, R. Das, M. Kumar and S. Mishra, Eur. Phys. J. E {\bf 42}, 62 (2019).

\bibitem{ref18} P. Malgaretti and H. Stark, J. Chem. Phys. {\bf 146}, 174901 (2017).

\bibitem{ref15} U. Choudhury, A.V. Straube, P. Fischer, J.G. Gibbs and F. Hofling, New J. Phys. {\bf 19}, 125010 (2017).

\bibitem{ref16} M.N. Popescu, W.E. Uspal, C. Bechinger and P. Fischer, Nano Lett. {\bf 18} 5345 (2018).

\bibitem{ref19} B.-q. Ai, Q.-y. Chen, Y.-f. He, F.-g. Li and W.-r. Zhong, Phys. Rev. E. {\bf 88}, 062129 (2013).

\bibitem{ref20} Cs. Sandor, A. Libal, C. Reichhardt and C.J.O. Reichhardt, Phys. Rev. E {\bf 95}, 012607 (2017).

\bibitem{ref21} C. Reichhardt and C.J.O. Reichhardt, Phys. Rev. E {\bf 97}, 052613 (2018).

\bibitem{ref22stark} A. Zottl and H. Stark, J. Phys.: Condens. Matter {\bf 28}, 253001 (2016).

\bibitem{ref12} J. Tailleur and M.E. Cates, Phys. Rev. Lett. {\bf 100}, 218103 (2008).

\bibitem{ref13} M.E. Cates and J. Tailleur, Annu. Rev. Condens. Matter Phys. {\bf 6}, 219 (2015).

\bibitem{ref14}  M.E. Cates and J. Tailleur, EPL {\bf 101}, 2 (2013).

\bibitem{thermo} A.P. Solon, J. Stenhammar, M.E. Cates, Y. Kafri and J. Tailleur, New J. Phys. {\bf 20}, 075001 (2018).

\bibitem{ref9} R. Wittkowski, A. Tiribocchi, J. Stenhammar, R.J. Allen, D. Marenduzzo and M.E. Cates, Nature Comm. {\bf 5}, 4351 (2014).
	
\bibitem{pw09} S. Puri and V. Wadhawan (eds.), {\it Kinetics of Phase Transitions}, CRC Press, Florida (2009).

\bibitem{AMB} S. Pattanayak, S. Mishra and S. Puri arXiv:2101.10626 (2021).

\bibitem{kardar} M. Kardar, G. Parisi and Y.-C. Zhang, Phys. Rev. Lett. {\bf 56}, 9 (1986).

\bibitem{dean} D.S. Dean, J. Phys. A: Math. Gen. {\bf 29}, L613 (1996).

\bibitem{pbl97} S. Puri, A.J. Bray and J.L. Lebowitz, Phys. Rev. E {\bf 56}, 758 (1997).

\bibitem{gbp05} S. van Gemmert, G.T. Barkema and S. Puri, Phys. Rev. E {\bf 72}, 046131 (2005).

\bibitem{GNF1} S. Ramaswamy, A.R. Simha and J. Toner, EPL {\bf 62}, 196 (2003).

\bibitem{ref30} D. Das, D. Das and A. Prasad, J. Theor. Biol. {\bf 308}, 96 (2012).

\bibitem{ref28} G. Gregoire and H. Chate, Phys. Rev. Lett. {\bf 92}, 025702 (2004).

\bibitem{GNF5} S. Pattanayak and S. Mishra, J. Phys. Commun. {\bf 2}, 045007 (2018).

\bibitem{po88} S. Puri and Y. Oono, J. Phys. A {\bf 21}, L755 (1988).

\bibitem{Chate} X.-q. Shi, G. Fausti, H. Chate, C. Nardini and A. Solon, Phys. Rev. Lett. {\bf 125}, 168001 (2020).

\bibitem{ref43} V. Banerjee, S. Puri and G.P. Shrivastav, Ind. J. Phys. {\bf 88}, 1005 (2014).

\bibitem{ref44} G.P. Shrivastav, V. Banerjee and S. Puri, Eur. Phys. J. E {\bf 37}, 1 (2014).

\end{thebibliography}
\end{document}